\documentclass[11pt]{article}
\usepackage{graphicx} 
\usepackage{amssymb,amsmath,amsthm} 
\usepackage{fullpage}
\usepackage{multirow}
\providecommand{\openone}{\leavevmode\hbox{\small1\kern-3.8pt\normalsize1}} 

\providecommand{\ket}[1]{|#1\rangle}

\providecommand{\ketbra}[2]{|#1\rangle\kern-2.8pt\langle#2|}

\newtheorem*{theorem*}{Theorem}
\newtheorem*{proposition*}{Proposition}

\def\be{\begin{equation}}
\def\ee{\end{equation}}

\newenvironment{narrow}[2]{ 
\begin{list}{}{
\setlength{\topsep}{0pt}
\setlength{\leftmargin}{#1}
\setlength{\rightmargin}{#2}
\setlength{\listparindent}{\parindent}
\setlength{\itemindent}{\parindent}
\setlength{\parsep}{\parskip}}
\item[]}{\end{list}}

\begin{document}
\title{Looking for symmetric Bell inequalities}
\author{Jean-Daniel Bancal$^1$, Nicolas Gisin$^1$, Stefano Pironio$^2$ \\[0.5em]
{\it $^1$Group of Applied Physics, University of Geneva,} \\
{\it 20 rue de l'Ecole-de-M\'edecine, CH-1211 Geneva 4, Switzerland}\\
{\it $^2$Laboratoire d'Information Quantique, Universit\'e Libre de Bruxelles, Belgium}}
\date{\today}
\maketitle

\begin{abstract}
Finding all Bell inequalities for a given number of parties, measurement settings, and measurement outcomes is in general a computationally hard task. We show that all Bell inequalities which are symmetric under the exchange of parties can be found by examining a symmetrized polytope which is simpler than the full Bell polytope. As an illustration of our method, we generate $238885$ new Bell inequalities and $1085$ new Svetlichny inequalities. We find, in particular, facet inequalities for Bell experiments involving two parties and two measurement settings that are not of the Collins-Gisin-Linden-Massar-Popescu type.
\end{abstract}

\section{Introduction}
Already discovered by Boole in the theory of logic and probabilities as ``conditions of possible experience'' \cite{Pitowsky}, Bell inequalities found a new dimension with the work of John Bell who showed that quantum physics could violate these conditions in some situations, highlighting what is now known as quantum nonlocality  \cite{Bell}.

Complete set of Bell inequalities are known only for setups involving small numbers of parties, measurement settings, and measurement outcomes. This may already be sufficient for various applications, such as exhibiting the nonlocality of a noisy quantum state in a real experiment \cite{Aspect}, or establishing the security of a device-independent quantum key distribution protocol \cite{Pironio09,masanes}. But the simplest inequalities are not always optimal. For instance, certain inequalities with large number of measurement settings are much more resistant to the detection inefficiencies than the CHSH inequality \cite{Vertesi09,Brunner}, or are violated by quantum states that do not violate the CHSH inequality \cite{Vertesi,Collins}. This motivates a search for Bell inequalities involving more parties, measurements, or outcomes. But finding all Bell inequalities pertaining to a given experimental setup is in general a hard task \cite{Pitowski,Werner}, and a complete search with current-day techniques is not feasible in most instances.

It is instructive, however, to realize that many useful Bell inequalities, like CHSH \cite{CHSH}, Mermin \cite{Mermin} or CGLMP \cite{CGLMP} to cite just a few, can be written in a form that is invariant under any permutation of the parties. Symmetric inequalities are also attractive because they are likely to be easier to handle. Motivated by this observation, we show here how to exploit a symmetric version of the full Bell polytope to generate all symmetric Bell inequalities. This symmetrized polytope is much easier to handle than the full Bell polytope. In particular, for Bell experiment with binary settings and outcomes, the number of extremal points of the symmetrized polytope only grows polynomially with the number of parties, which is an exponential gain compared to the general situation. 

Our method for finding symmetric inequalities is not restricted to Bell inequalities, but can be applied to any set of inequalities characterizing a correlation polytope, like for instance the Svetlichny inequalities, which allow to test for genuinely multipartite nonlocality~\cite{Svetlichny}. We present in the next section our approach from this general perspective. We then apply in Section~3 our method to several examples for which listing all (non-symmetric) inequalities is computationally intractable with present-day techniques. In the Bell scenario with two parties, two settings, and four outcomes, we find in particular facet inequalities that are not of the CGLMP form, answering an open question raised by Gill~\cite{gill}.

\section{General setting}
Let ($n$, $m$, $k$) denote a Bell experiment where $n$ parties can choose one out of $m$ possible measurement settings that each yield one out of $k$ possible outcomes\footnote{More general Bell scenarios with different number of measurement settings $m_i$ and outcomes $k_i$ for each party~$i$ are also possible, but since we will consider situations that are symmetric under permutations of the parties, we choose $m_i=m$, $k_i=k$ for all $i$.}. The statistics of the observed results are described by the joint conditional probability distributions (also called correlations\footnote{Throughout this paper, the term ``correlations" refers to probability distributions of the form \eqref{eq:P}. It should not to be confused with ``correlator" or ``correlation functions", such as $E(s_1,s_2)=p(r_1=r_2|s_1,s_2)-p(r_1\neq r_2|s_1,s_2)$.})
\begin{equation}\label{eq:P}
p(r_1,\ldots,r_n|s_1,\ldots,s_n)
\end{equation}
where $s_i\in\{1,\ldots,m\}$ denotes the measurement setting of party $i$ and $r_i\in\{1,\ldots,k\}$ denotes the corresponding measurement result.

Note that in general the $N=m^nk^n$ probabilities (\ref{eq:P}) are not all independent but satisfy linear constraints, such as the normalization conditions or the no-signalling conditions. We are thus actually interested in an affine subspace of $\mathbb{R}^N$ of dimension $d$, where $d=m^n(k^n-1)$ for normalized correlations, and $d=(m(k-1)+1)^n-1$ for correlations that satisfy in addition the no-signalling conditions \cite{Collins,liftings}. We suppose in the following that a proper parametrization has been introduced so that the joint distributions (\ref{eq:P}) can be identified with points $p$ in $\mathbb{R}^d$.

We are interested in whether a given $p$ belongs to some special subset $P\subseteq \mathbb{R}^d$.  In this work, we consider sets $P$ which are polytopes\footnote{In general, it might also be interesting to consider sets that are not polytopes~\cite{npa,npa2}, nor even convex~\cite{Branciard}.}. A polytope can be described by the list ${V}$ of its vertices (or extremal points) $v\in {V}$, and any point $p\in P$ can be written as
\begin{equation}
p=\sum_{v\in {V}} \rho_v v\,,
\end{equation}
where $\rho_v$ are positive and normalized weights: $\rho_v\geq 0$ for all $v$ and $\sum_{v\in {V}} \rho_v = 1$.
In the case of the Bell polytope, for instance, the extremal points are the deterministic local strategies, corresponding to the prior assignment of an outcome $r_{i,s_i,v}$ to each measurement setting $s_i$. There are thus $k^{nm}$ different vertices $v$, corresponding to joint probability distributions of the form
\be\label{eq:localstrategy}
p(r_1,\ldots,r_n|s_1,\ldots,s_n)=\begin{cases} 1&\text{if $r_i=r_{i,s_i,v}$  for all $i=1,\ldots,n$}\\
0&\text{otherwise}\,.
\end{cases}
\ee

Alternatively to its representation in term of vertices (the $V$-repesentation), a polytope $P$ can also be described, by the Farkas-Minkowski-Weyl theorem, as the intersection of finitely many half-spaces (the $H$-representation). A half-space is defined by an inequality $h\cdot p=\sum_{i=1}^d h_i\,p_i\leq h_0$ specified by the couple $(h, h_0)\in \mathbb{R}^{d+1}$. We say that an inequality $(h,h_0)$ is valid for the polytope $P$, if $h\cdot p\leq h_0$ for all points $p$ in $P$. The Farkas-Minkowski-Weyl theorem states that a polytope can be characterized by a finite set of valid inequalities. That is, a point $p$ belongs to $P$ if and only~if
\be\label{hs}
h\cdot p\leq  h_0\quad\text{ for all } ( h,h_0)\in H\,,
\ee
where $H$ is some finite set in $\mathbb{R}^{d+1}$.
This description is particularly appropriate when willing to show that a point does not belong to the polytope, since it is sufficient to exhibit the violation of a single one of the inequalities (\ref{hs}). In the case of the Bell polytope, these inequalities are known as Bell inequalities.

A complete and minimal representation of a polytope in the form (\ref{hs}) is provided by the set of facets of the polytope. An inequality $(f,f_0)\in H$ defines a facet if its associated hyper-plane $f\cdot p=f_0$ intersects the boundary of the polytope in a set of dimension $d-1$, i.e., if there exists $d$ affinely independent points
of $P$ satisfying $f\cdot p=f_0$. The set $\hat f=\{p\mid p\in P, \,f\cdot p=p_0\}$ then corresponds to the facet defined by $(f,f_0)$.

Finding all the Bell inequalities corresponding to an experimental configuration $(n,m,k)$ thus amounts to determining the facets (minimal H-representation) of a polytope when given its extremal points (V-representation). This conversion problem is well known and there exists several available algorithms to solve it \cite{polytopeSoft}. However, when $n$, $m$, or $k$ are too large, the associated polytope becomes too complex to be handled by these algorithms. It might then be appropriate to focus the search on a subclass of all facet inequalities. We explain how this can be done for symmetric inequalities in the next section.

\subsection{Focusing on symmetric facets}
Some facet inequalities $(f,f_0)$ of a correlation polytope $P$ can be written in a way that is invariant under permutations of the parties. This is the case for instance for the CHSH inequality, which can be written in the CH form as
\be\label{chsh}
p(a_1)+p(b_1)-p(a_1b_1)-p(a_1b_2)-p(a_2b_1)+p(a_2b_2)\geq 0
\ee
where we write $p(a,b|x,y)$ for $p(r_1,r_2|s_1,s_2)$\footnote{From now on, we write $p(a,b,c\ldots|x,y,z,\ldots)$ for  $p(r_1,r_2,r_3,\ldots|s_1,s_2,s_3,\ldots)$.} and define $p(a_x)=p(a=1|x)$, $p(b_y)=p(b=1|y)$, $p(a_xb_y)=p(a=1,b=1|x,y)$.
In the following we will call such facets \emph{symmetric facets}\footnote{Note that a symmetric inequality need not appear in a symmetric form when written in any of its equivalent forms under relabeling of measurement settings and outcomes. For instance : $p(a_1)+p(b_2)-p(a_1b_1)-p(a_1b_2)+p(a_2b_1)-p(a_2b_2)\geq 0$ is equivalent to the CH inequality (\ref{chsh}) up to relabeling of the settings and outcomes, but is not invariant under the exchange of party $1$ and $2$.}. If one is interested in finding only the symmetric Bell inequalities relevant to a given scenario $(n,m,k)$, then it is possible to restrict the space in which to search for them. This is what we show now.

Let $G$ be the set of permutations of $\{1,\ldots,n\}$. Given a permutation $\pi\in G$ let
\be
p(a_1,\ldots,a_n|x_1,\ldots,x_n)\mapsto p(a_{\pi(1)},\ldots a_{\pi(n)}|x_{\pi(1)},\ldots,x_{\pi(n)})\,.
\ee
be its action on the joint probability distributions. This permutation induces a transformation $\pi\,:\,\mathbb{R}^d\mapsto\mathbb{R}^d\,:\, p\mapsto \pi p$ in the vector space $\mathbb{R}^d$ in which the correlations $p$ are represented, which by abuse of language we denote by the same symbol~$\pi$. Note that the correlation polytopes that we consider here are evidently invariant under such permutations, i.e., $\pi P=P$, and any vertex $v\in V$ is mapped into another vertex $\pi v\in V$.

Given a facet-inequality $(f,f_0)$ of $P$, let $(\pi f,f_0)$ be its image under $\pi$. Note that with this definition the facet $\hat f=\{p\mid p\in P,\,f\cdot p=f_0\}$ is mapped onto $\pi\hat f=\{\pi p\mid p\in P,\,f\cdot p=f_0\}$ since $\{p\mid p\in P, (\pi f) \cdot p=f_0\}=\{p\mid p\in P, f \cdot (\pi^{\dagger} p)=f_0\}=\{\pi p\mid p\in P,\,  f\cdot p=f_0\}$, where we used the fact that the transformations $\pi$ are unitary, i.e., $\pi^\dagger=\pi^{-1}$.
 We say that a facet $(f,f_0)$ is \emph{symmetric} if $\pi f =f$ for all $\pi\in G$.

 Consider now the symmetrizing map
\begin{equation}
\Pi=\frac1{|G|}\sum_{\pi\in G}\pi
\end{equation}
and let $\tilde \Pi$ denote the projection on the symmetric affine subspace $S$ of $\mathbb{R}^d$ of dimension $d_s=\text{dim}(S)$. We suppose to simplify the presentation that the origin of $\mathbb{R}^d$ is contained in $S$ (this can always be achieved by a proper translation of the correlation vectors $p$), so that $S$ is actually a linear subspace of $\mathbb{R}^d$.
An arbitrary vector $p\in \mathbb{R}^d$ can then be written as $p=p_s+ p_{t}$ where $p_s=\tilde \Pi p$ is the projection of $p$ on the symmetric subspace $S$ and $p_t=(1-\tilde \Pi)p$ on the complementary space $T=
(1-\tilde \Pi)\mathbb{R}^d$.
Similarly an arbitrary inequality $(f,f_0)$ can be written as $(f_s\oplus f_t, f_0)$. A symmetric inequality then takes the form $(f_s\oplus 0,f_0)$ where ${0}$ denotes the null vector in $\mathbb{R}^{d-d_s}$.
Given an inequality $(f_s,f_{0})\in \mathbb{R}^{d_s+1}$ defined in the symmetric subspace $S$, we denote its \emph{symmetric extension} as the inequality $(f,f_0)=(f_s\oplus {0},f_{0})$ defined in the full space $\mathbb{R}^d$.

Let $P_s=\tilde \Pi P=\{p_s\mid p\in P\}$ be the projection of the polytope $P$ on the symmetric subspace.
Any vertex $w\in P_s$ is necessarily the projection $w=v_s$ of some vertex $v\in P$. Indeed, suppose that $w$ is the projection of a non-extremal point $p=\sum_iq_iv_i\in P$, $\sum_i q_i=1$, $0<q_i<1$. Then it needs to be the projection of every $v_i$ as well: $w=\tilde\Pi p =\sum_iq_i\tilde\Pi v_i = w$ implies $w=\tilde\Pi v_i$ since $w$ is a vertex. Note, however, that every vertex $v\in P$ does not necessarily induce a vertex $v_s\in P_s$ when projected on the symmetric subspace. The symmetrized polytope $P_s$ has thus in general less vertices than the original polytope. Moreover it is defined in a space of smaller dimension $d_s<d$ than the full space $\mathbb{R}^d$. It is thus in general easier to determine the facets of the symmetrized polytope $P_s$ than those of the full polytope $P$. The following theorem shows that determining the facets of this symmetrized polytope is sufficient to find all symmetric facet inequalities of the full polytope $P$ (see also Figure~1).
\begin{theorem*}\label{thm}
Let $(f_s,f_{0})$ be a facet inequality for the polytope $P_s\in\mathbb{R}^{d_s}$. Then its symmetric extension $(f_s\oplus 0,f_0)\in\mathbb{R}^{d+1}$ defines a valid inequality for the full polytope $P\in\mathbb{R}^d$. Moreover all symmetric facet inequalities of the full polytope $P$ are the symmetric extension of some facet of the symmetrized polytope $P_s$.
\end{theorem*}
\begin{proof}
The symmetric extension $(f,f_0)=(f_s\oplus 0,f_0)$ is valid for $P$ if $f\cdot p\leq f_0$ is satisfied by all points $p\in P$. But this immediately follows from the fact that $f\cdot p = ( f_s\oplus {0})\cdot (p_s\oplus p_t) =  f_s \cdot p_s\leq f_{0}$ and the fact that $f_s\cdot p_s\leq f_0$ is valid for all $p_s\in P_s$.

Now, let $(g,g_0)=(g_s\oplus 0, g_0)$ be a symmetric facet of $P$. Clearly, it is the symmetric extension of the inequality $(g_s,g_0)\in\mathbb{R}^{d_s}$, which is valid for $P_s$. Moreover, $(g_s,g_0)$ defines a facet of $P_s$ as there exist $d_s$ affinely independent points in $P_s$ that saturate it. Indeed, since $(g, g_0)$ defines a facet of $P$, there exists $d$ affinely independent points $p$ in $P$ that satisfy $(g_s\oplus 0)\cdot p=g_0$. These points are of the form $p=p_s\oplus p_t$, where $p_t$ can clearly be arbitrary. Since the complementary space $T$ is of dimension $d-d_s$, there must therefore be at least $d-(d-d_s)=d_s$ affinely independent points of the form $p_s\oplus 0$ that saturate the inequality $(g_s\oplus 0,g_0)$. These points obviously define $d_s$ affinely independent points in $P_s$ that saturate the inequality $(g_s,g_0)$.
\end{proof}
\begin{figure}
\begin{center}
\includegraphics[width=13cm]{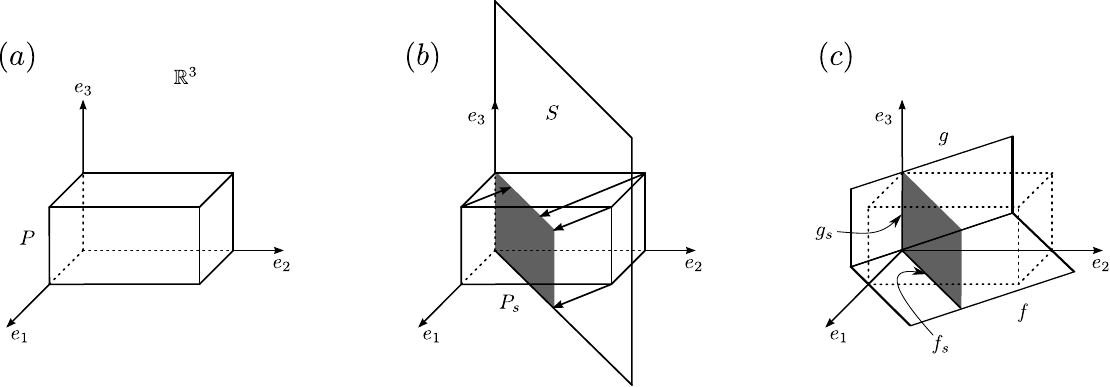}
\caption{(a) Example of a polytope $P$ in the vector space $\mathbb{R}^3$. (b) Subspace $S$ symmetric under the exchange of coordinates $e_1$ and $e_2$. $P_s$ (grey) is the projection of the polytope onto this subspace. (c) $f_s$ and $g_s$ are two facets of $P_s$, and $f$ and $g$ are their symmetric extensions to the whole space $\mathbb{R}^3$. $f$ is a symmetric facet of the original polytope $P$, whereas $g$ is just a valid inequality for $P$.}
\label{fig:fig1}
\end{center}
\end{figure}
Note however that the converse of the theorem is not true, as illustrated in Figure~\ref{fig:fig1}: facets of $P_s$ do not necessarily extend to facets of the polytope $P$ in the general space $\mathbb{R}^d$. We show in Section 2.3 how it is nevertheless possible to take advantage of such inequalities to generate new (not necessarily symmetric) facet inequalities for the original polytope $P$.

\subsection{Illustration on the $(2,2,2)$ Bell scenario}
We now illustrate in detail the above approach on the $(2,2,2)$ Bell scenario. This scenario is characterized by 16 probabilities $p(ab|xy)$, where $x\in\{1,2\}$ denote the measurement setting of Alice and $a\in\{1,2\}$ the corresponding outcome, and where similarly $y\in\{1,2\}$ and $b\in\{1,2\}$ denote Bob's measurement setting and outcome. These probabilities satisfy normalization
\be
\sum_{a,b=1,2} p(ab|xy)=1\quad\text{ for all } x,y=1,2,
\ee
and no-signalling
\begin{eqnarray}
&&p(a|x)\equiv\sum_{b=1,2} p(ab|xy)\quad \text{ for all } a,x,y=1,2\nonumber\\
&&p(b|y)\equiv\sum_{a=1,2} p(ab|xy)\quad \text{ for all } b,x,y=1,2\,.
\end{eqnarray}
In total, only $8$ of the 16 probabilities $p(ab|xy)$ are therefore independent and we can represent the correlations $p$ as elements of $\mathbb{R}^8$. For specificity, we choose the following parametrization
\begin{equation}\label{eq:nosigparam}
p=\left[p(a_1),p(a_2),p(b_1),p(b_2),p(a_1b_1),p(a_1b_2),p(a_2b_1),p(a_2b_2)\right]\,,
\end{equation}
where $p(a_x)=p(1|x)$, $p(b_y)=p(1|y)$ and $p(a_xb_y)=p(11|xy)$.

The Bell-local polytope is described by 16 vertices
\be\label{det222}
v_{a_1a_2b_1b_2}=\left[a_1,a_2,b_1,b_2,a_1b_1,a_1b_2,a_2b_1,a_2b_2\right]\,,
\ee
where $a_1,a_2,b_1,b_2\in\{1,2\}$ specifies the deterministic assignment of an outcome to each measurement setting.

The group $G$ of permutation of two parties contains 2 elements: the identity $\openone$ and the permutation $\pi$ acting as follows on a vector $p$,
\begin{equation}
\pi p=\left[p(b_1),p(b_2),p(a_1),p(a_2),p(a_1b_1),p(a_2b_1),p(a_1b_1),p(a_2b_2)\right]\,.
\end{equation}
The symmetrizing map projecting on the space $S$ of symmetric correlations is thus equal to $\Pi=\frac12(\openone+\pi)$, while the map projecting on the complementary space $T$ is $\openone-\Pi=\frac12(\openone-\pi)$. Arbitrary correlations $p$ can thus be decomposed into a symmetric and an asymmetric part
$
p=p_s\oplus p_t,
$
where
\begin{eqnarray}
p_s &=& \frac{p+\pi p}{2}\nonumber \\
&=&\left[\frac{p(a_1)+p(b_1)}{2}, \frac{p(a_2)+p(b_2)}{2}, \frac{p(a_1)+p(b_1)}{2}, \frac{p(a_2)+p(b_2)}{2},\right. \\ &&\left.\quad p(a_1b_1),\frac{p(a_1b_2)+p(a_2b_1)}{2},\frac{p(a_1b_2)+p(a_2b_1)}{2},p(a_2b_2)\right]\,,\nonumber
\end{eqnarray}
and
\begin{eqnarray}
p_t &=& \frac{p-\pi p}{2}\nonumber \\
&=&\left[\frac{p(a_1)-p(b_1)}{2}, \frac{p(a_2)-p(b_2)}{2}, \frac{-p(a_1)+p(b_1)}{2}, \frac{-p(a_2)+p(b_2)}{2},\right.\\ &&\quad \left.0,\frac{p(a_1b_2)-p(a_2b_1)}{2},\frac{-p(a_1b_2)+p(a_2b_1)}{2},0\right]\,.\nonumber
\end{eqnarray}
Note that the symmetric part $p_s$ lives in a 5-dimensional subspace of $\mathbb{R}^8$ and can thus be expressed in an appropriate basis as
\be\label{sym5}
p_s = \left[\frac{p(a_1)+p(b_1)}{2}, \frac{p(a_2)+p(b_2)}{2},p(a_1b_1),\frac{p(a_1b_2)+p(a_2b_1)}{2},p(a_2b_2)\right]
\ee
Similarly, $p_t$ lives in a 3-dimensional space of $\mathbb{R}^8$ and can be decomposed in a proper basis as
\be
p_t = \left[\frac{p(a_1)-p(b_1)}{2}, \frac{p(a_2)-p(b_2)}{2},\frac{p(a_1b_2)-p(a_2b_1)}{2}\right]\,.
\ee

The projection of the 16 deterministic points (\ref{det222}) on the symmetric subspace defined by (\ref{sym5}) are given by
\be\label{dets222}
v_{s;a_1a_2b_1b_2}=\left[\frac{a_1+b_1}{2},\frac{a_2+b_2}{2},a_1b_1,\frac{a_1b_2+a_2b_1}{2},a_2b_2\right]\,.
\ee
Note that some vertices of the original polytope are projected onto the same point of the symmetric polytope. For instance, $v_{s;1112}=v_{s;1211}$. In total, it can be verified that the set defined by (\ref{dets222}) contains only 10 extremal points.

We thus have reduced the original 8-dimensional polytope defined by 16 vertices to a 5-dimensional polytope with 10 vertices. Applying to this symmetrized polytope a standard algorithm performing the transformation from the V-representation to the H-representation, we find 4 different types (up to relabeling of settings and outcomes) of facets of the symmetric polytope:
\begin{align}
&P(a_1b_1)\geq0&\nonumber\\
&P(a_1b_2)+P(a_2b_1)\geq0&\nonumber\\
&P(a_1)+P(b_1)-2P(a_1b_1)\geq0&\\
&P(a_1)+P(b_1)-P(a_1b_1)-P(a_1b_2)-P(a_2b_1)+P(a_2b_2)\geq0&\nonumber
\end{align}
We recognize the first inequality as the positivity condition for the joint probabilities and the last one as the CHSH inequality (written in the CH form as in equation \eqref{chsh}). These two classes of inequalities define symmetric facets of the full polytope. The two other inequalities are valid inequalities for the full local polytope, but do not correspond to facets (although they are facets of the symmetrized polytope, as the inequality $g$ in Figure~\ref{fig:fig1}).
Note that in this (2,2,2) Bell scenario, the two only types of facet inequalities of the full polytope (the positivity condition and the CHSH inequality) can be written in a symmetric way. Hence in this simple case finding the facets of the symmetrized polytope is sufficient to generate all Bell inequalities.

\subsection{Generating facet inequalities from valid inequalities}\label{sousIne}
As we mentioned earlier, and was illustrated above, facets of the symmetrized polytope $P_s$ can correspond to inequalities which are not facets of the original polytope $P$. These inequalities are nonetheless valid inequalities which are satisfied by all points in $P$ and which might be violated by points that do not violate any of the symmetric facets of $P$. These inequalities may be used to generate new (non-symmetric) facets of $P$.

There exist various deterministic or heuristic algorithms which may generate new facet inequalities starting from a valid (not facet-defining) inequality, see for instance \cite{Pal}. Here, we describe a procedure that can be used whenever the starting valid inequality corresponds to a high-dimensional face of $P$, i.e., when the number of affinely independent vertices that saturate the inequality is large. In this case, it is possible to find all the facets that contain this high-dimensional face by completing the list of vertices with all possible combination of vertices that do not saturate the inequality, as detailed by the following algorithm:

\begin{enumerate}
\item Let $(f,f_0)$ be a valid inequality for $P$ and let $W=\{v\mid f\cdot v=f_0\}$ be the set of vertices saturating this inequality.
\item Let $\text{dim}(W)$ denote the number of affinely independent points in $W$.
\begin{itemize}
\item If $\text{dim}(W)=d$, then $(f,f_0)$ is a facet of $P$.
\item If $\text{dim}(W)<d$, let $U=\{v\mid \text{dim}(W\cup v)>\text{dim}(W)\}$.\\
For every $u\in U$, let $(g,g_0)$ be the hyperplane passing through the points in $U$. If $(g,g_0)$ is a valid inequality for $P$, it now defines a face of $P$ of dimension $\text{dim}(W)+1$; in this case, go back to point 1 with $(g,g_0)$ as a starting inequality.
\end{itemize}
\end{enumerate}

\section{Applications}
We now illustrate our method in several situations for which generating the complete set of facet inequalities using standard polytope software \cite{polytopeSoft} is too time-consuming to be feasible. Due to the large number of inequalities that we have found, we only explicitly write a few of them here. Complete lists of all the facet inequalities that we generated are posted on the website \cite{website}. Our results are summarized in Table I.

Note that we list here only inequalities that belong to different equivalence classes, where two inequalities are considered equivalent if they are related by a relabeling of parties, measurement settings, or measurement outputs, or if they correspond to two different liftings of the same lower-dimensional inequality \cite{liftings}. In appendix A, we introduce a parametrization of the correlation space that naturally induces several invariants for each equivalence class. These invariants are easily computed and are useful to determine quickly wether two inequalities are equivalent (two equivalent inequalities have equal invariants).

\begin{table}[b]\label{tab:table}
\begin{tabular}{|r|c|c|c|c|c|c|}
\hline {Bell scenario} & {$d$}  & {$d_s$} &  {$|V|$} & {$|V_s|$} & \# symmetric inequalities & \# symmetric facets \\
\hline
(2,2,4)&48&27&256&136&29&12\\
(2,4,2)&24&14&256&136&90&55\\
(4,2,2)& 80 & 14& 256  & 35 & 627 & 392\\
(5,2,2)& 242 & 20 & 1024 & 56 & $>$238464 & 238464\\
Correlators (3,3,2)& 27 & 10 & 512 & 40 & 44 & 20\\
Svetlichny (3,2,2)& 56 & 14& 2944  & 132 & 1204 & 1087\\\hline
\end{tabular}
\caption{Summary of our numerical results. For each scenario we give the dimension of the space the polytope lives in as well as its number of extremal vertices, both before ($d$, $|V|$) and after projection on the symmetric subspace ($d_s,|V_s|$). We give the number of inequivalent symmetric inequalities valid for $P$ (but not necessarily facet-defining) obtained by resolving the symmetric polytope $P_s$ and the number of those inequalities that are facet defining.}
\end{table}

\subsection{(2,2,4)}
We first consider bipartite experiments with two measurement settings per sites and $4$ possible outcomes. Note that the case $(2,2,3)$ was completely solved by Kaszlikowski et al. \cite{KKCZO} and Collins et al. \cite{CGLMP}, who showed that all facets of the (2,2,3) polytope either correspond to the positivity of probabilities, the CHSH inequality, or the CGLMP inequality \cite{CGLMP}. The CGLMP inequality was introduced for any number of outcomes $k\geq 3$ in \cite{CGLMP}, and Gill raised the question \cite{gill} whether all non-trivial facet inequalities of $(2,2,k)$ are of this form.

Using our method, we found that the Bell polytope corresponding to $(2,2,4)$ contains 12 inequivalent symmetric Bell inequalities. Among them, 8 involve the four possible outcomes in a nontrivial way, i.e., they correspond to genuine 4-outcome inequalities that cannot be seen as liftings of inequalities with lower numbers of outcomes, these inequalities are listed in Appendix B. The list of these 8 inequalities contains the CGLMP inequality, but surprisingly, it also contains 7 inequalities that are inequivalent to it, thus answering in the negative Gill's question~\cite{gill}.

\subsection{(2,4,2)}
We now consider a bipartite scenario involving 4 settings with binary outcomes. The simpler case (2,3,2) was solved in \cite{CHSH,Froissart,Collins} and contains a single new inequality besides the positivity constraints and the CHSH inequality.

With $4$ settings, we could use our method to find all of the 90 inequivalent facets of the symmetrized polytope. Among these, 55 turn out to be facets of the (2,4,2) full local polytope, there are thus in total 55 symmetric inequalities for this scenario. Most of them were already known (see \cite{Brunner, Pal, Avis}), but we could not find the two following ones in the litterature, given here in the notation of \cite{Collins} :

 \begin{narrow}{-0.8cm}{0cm}
\begin{equation}
S_{(2,4,2)}^{51}=\begin{array}{c||cccc}
  & -1 & -2 & -2 & -2\\
\hline \hline
-1 &-3 & 3 & 2 & 2\\
-2 & 3 & 2 &-1 &-1\\
-2 & 2 &-1 &-1 & 3\\
-2 & 2 &-1 & 3 & 0
\end{array} \leq 0\ , \ \
S_{(2,4,2)}^{52}=\begin{array}{c||cccc}
  & 0 & -2 & -2 & -2\\
\hline \hline
 0 &-3 & 2 &-2 & 1\\
-2 & 2 & 0 & 2 & 2\\
-2 &-2 & 2 & 4 &-1\\
-2 & 1 & 2 &-1 & 1
\end{array} \leq 0
\end{equation}
 \end{narrow}



\subsection{(4,2,2) and (5,2,2)}
Since our method takes advantage of the symmetry between parties, we expect that it will be particularly useful for multipartite Bell scenarios. Indeed for the Bell scenario (n,2,2), corresponding to $n$ parties with binary settings and outcomes, the full local polytope has $4^n$ vertices and is embedded in a space of dimension $3^n-1$. The symmetrized polytope, on the other hand, has at most $\frac16(n+1)(n+2)(n+3)$ vertices and is embedded in a space of dimension $d_s=\frac12n(n+3)$. These quantities are polynomial in $n$ and represent an exponential advantage with respect to the general, non-symmetric situation. Note that it therefore follows that it is possible using linear programming to decide in polynomial time in $n$ if a given symmetric correlation vector $p$ is local.

The case (3,2,2) was already completely solved in  \cite{Sliwa}. For (4,2,2) we found a total of 627 inequivalent symmetric inequalities, of which 392 are facet-defining. These facet inequalities correspond to the positivity conditions and to 391 genuinely 4-partite inequalities. Amongst them, the following one is quite interesting, as it can be violated by a 4-partite W state with measurements lying in the x-y plane (contrary to the 3-partite case where no inequality is known that can be violated by a $W$-state with measurements lying in the x-y plane):
\begin{equation}
\begin{split}\label{4partin}
I_W =& - p(a_1b_1) + p(a_1b_1c_1) + p(a_1b_1c_2) - p(a_2b_2c_2) - p(a_1b_1c_1d_1) \\
&- p(a_1b_1c_1d_2) - p(a_1b_1c_2d_2) + p(a_1b_2c_2d_2) + p(a_2b_2c_2d_2) + \text{sym} \leq 0.\\
\end{split}
\end{equation}
The notation ``$\text{sym}$'' stands for the symmetric terms that are missing in (\ref{4partin}), such as $p(a_1c_1),p(a_1b_2c_1)$, etc. If we consider the $W$ state $|0001\rangle+|0010\rangle+|0100\rangle+|1000\rangle$ and measurements in the $x-y$ plane at an angle $\phi$ with respect to the $x$ axis, and set
\begin{align}
\phi_{A_1} &= \phi_{C_1} = 0 & \phi_{B_1}&=\phi_{D_1}=\arccos\frac14-2\arcsin\frac14\nonumber\\
\phi_{A_2} &= \phi_{C_2} = \arccos\frac14 & \phi_{B_2}&=\phi_{D_2}=-2\arcsin\frac14
\end{align}
we find a value $I_W=1/16>0$.

For (5,2,2), we found 238464 inequivalent symmetric facets. Note that all of these inequalities, except the positivity of the probabilities, are truly 5-partite ones (i.e., they do not correspond to lifting of inequalities involving less parties). Among these inequalities, 9 of them involve only full (5-partite) correlators and were already given in \cite{Werner}.

\subsection{Correlation inequalities for (3,3,2)}
We considered also a tripartite scenario with 3 binary measurements per party. Since in this case even the symmetrized polytope is quite time-consuming to solve, we made a further restriction by considering only ``full-correlator" inequalities, which can be written using only terms of the form $\langle A_xB_yC_z\rangle=p(a+b+c=0|x,y,z)-p(a+b+c=1|x,y,z)$. This corresponds to performing a projection of the polytope on the subspace defined by
\begin{equation}
\langle A_i\rangle=\langle B_i\rangle=\langle C_i\rangle=\langle A_iB_j\rangle=\langle A_iC_j\rangle=\langle B_iC_j\rangle=0\ \quad \text{for all } i,j=1,2,3.
\end{equation}
We obtained 40 inequalities in this way, 18 of which are facets of the full original polytope that truly involve 3 inputs per party. Using the method presented in section \ref{sousIne}, we found 13 supplementary facet inequalities, all of which involve again full-correlators only.

\subsection{Svetlichny inequalities for (3,2,2)}
To illustrate that our method is not restricted to Bell-local polytope, but can address any correlation polytope, we consider the Svetlichny polytope for 3 parties \cite{Svetlichny}, which characterizes true tripartite nonlocality \cite{Seevinck,Collins2,Bancal}.

In a Svetlichny model, two of the three subsystems are allowed to communicate once the measurement settings have been chosen. There are thus three types of Svetlichny vertices $v_{AB/C}$, $v_{AC/B}$, $v_{BC/A}$, depending on which pairs of parties are allowed to communicate. A vertex of the form, e.g. $v_{AB/C}$, corresponding to a deterministic strategy where outcomes $\alpha(x,y)$ and $\beta(x,y)$ are jointly determined for party 1 and 2, and an outcome $\gamma(z)$ is assigned to party 3. This defines a joint distribution of the form
\be\label{eq:sverstrategy}
p(a,b,c|x,y,z)=\begin{cases} 1&\text{if $a=\alpha(x,y), b=\beta(x,y), c=\gamma(z)$}\\
0&\text{otherwise}\,.
\end{cases}
\ee
Such probability points do not satisfy the no-signalling conditions. For binary settings and outcomes, the Svetlichny polytope thus lives in a vector space of dimension $d=56$. The subspace which is symmetric under the exchange of the three parties, however, has only dimension $d_s=14$.

This great reduction in the space dimension, together with a reduction in the number of extremal points (see Table 1), allowed us to find all symmetric Svetlichny inequalities. After projection on the no-signaling space\footnote{Indeed, we are only interested in whether these inequalities are violated by quantum correlations, which satisfy the no-signalling conditions.} a total of 1087 facet symmetric Svetlichny inequalities were found.

Interestingly, there are only two symmetric Svetlichny inequalities that involve only full-triparite correlation terms: the original Svetlichny inequality \cite{Svetlichny} and the following one:
\begin{equation}
\begin{split}
I_{Corr}=&- \langle A_1 B_1 C_1\rangle + \langle A_1 B_1 C_2\rangle + \langle A_1 B_2 C_1\rangle - 3\langle A_1 B_2 C_2\rangle\\
&+ \langle A_2 B_1 C_1\rangle - 3\langle A_2 B_1 C_2\rangle - 3\langle A_2 B_2 C_1\rangle - 3\langle A_2 B_2 C_2\rangle \leq 10
\end{split}
\end{equation}
This last inequality can be violated by quantum states, for instance by GHZ states having a visibility larger than $95.68\%$.

The following inequality is also interesting:
\begin{equation}
I_{GHZ}=-3P(a_2)+P(a_1b_2)-P(a_1b_1c_1)-P(a_1b_2c_2)+7P(a_2b_2c_2)+\text{sym} \leq 0.
\end{equation}
where ``$\text{sym}$'' stands for the missing symmetric terms. It can be shown \cite{Svet10} that it is violated by every GHZ-like state of the form
\begin{equation}\label{eq:GHZ}
\ket{GHZ}=\cos\theta\ket{000}+\sin\theta\ket{111}.
\end{equation}

\section{Outlook}
Motivated by the number of interesting Bell inequalities that are invariant under permutations of the parties, we introduced a method to list all symmetric inequalities. This method works even in cases where solving the full correlation polytope is computationally intractable. Our method can also be used as a starting point to generate more general, non-symmetric inequalities using algorithms such as the one described in section \ref{sousIne} or in \cite{Pal}.

Our method allowed us to find a number of new Bell inequalities. But a new problem is now at sight: so many different inequalities are generated, even for simple situations, that it is difficult to find which ones are the most interesting. Evidence of this problem was already put forward in \cite{Werner} and in \cite{Avis} where it was shown that the number of inequivalent Bell inequalities increases very quickly with the number of parties or measurement settings. New insights are thus necessary in order to classify these inequalities and understand which ones are the most relevant.

For simple Bell scenarios, such as (2,2,2), (2,2,3) or (2,3,2) \emph{all} facet inequalities happen to be symmetric inequalities. What is the proportion of symmetric inequalities in more complicated scenarios? Are there other useful symmetries or properties that can be exploited to generate more inequalities?

\section{Acknowledgments}
We thank K. F. P\'al and T. V\'ertesi for useful remarks. This work was supported by the Swiss NCCR Quantum Photonics, the European ERC-AG QORE, and the Brussels-Capital region through a BB2B grant.

\appendix
\section*{Appendix A: Correlators and inequality invariants}
When dealing with Bell inequalities, it is useful to check quickly if two inequalities are equivalent under relabeling of parties, measurement settings, or outcomes. We were confronted with this problem when classifying the inequalities that we derived here. In this appendix, we introduce a parametrization of the correlation space which naturally induces several invariants for each equivalence class. Two inequalities that are equivalent have equal invariants.

Let us consider a $(n,m,k)$ Bell scenario and let $I\subseteq\{1,\ldots,n\}$ be a subset of the $n$ parties. Define $p(a_I|x)$ as the probability that these $|I|$ parties obtain outputs $a_I=a_{i_1},\ldots,a_{i_{|I|}}$ given that measurements $x=x_1,\ldots,x_n$ have been made on all parties. Note that with this notation, the no-signalling condition is expressed as $p(a_I|x)=p(a_I|x_I)$, where $x_I=x_{i_1},\ldots,x_{i_{|I|}}$.

Define now single-party ``correlators" $E(a_i|x)$ by
\begin{equation}\label{eq:E}
E(a_i|x)=kp(a_i|x)-1\,,
\end{equation}
and define by induction multipartite correlators $E(a_I|x)$ through
\begin{equation}\label{eq:E2}
E(a_{I_1}, a_{I_2} | x) = E(a_{I_1}|x)*E(a_{I_2}|x)\,,
\end{equation}
where the $\ast$ operation is just the usual multiplication, except when acting on two probabilities, in which case it satisfies $p(a_{I_1}|x)*p(a_{I_2}|x)=p(a_{I_1},a_{I_2}|x)$.
The no-signalling condition is then expressed as
\be
E(a_I|x)=E(a_I|x_I)\,.
\ee
The normalization condition on the probabilities $p(a_I|x)$ imply, on the other hand, that for any $i\in I$
\begin{equation}\label{eq:normalisation}
\begin{split}
\sum_{a_i=0}^{k-1}E(a_I|x)&=\sum_{a_i=0}^{k-1}E(a_i|x)*E(a_{I\backslash i}|x)\\
&=\left[\sum_{a_i=0}^{k-1}\left(kp(a_i|x)-1\right)\right]*E(a_{I\backslash i}|x)=0\,.
\end{split}
\end{equation}

The correlators $E(a_I|x)$ are in one-to-one correspondence with the probabilities $p(a_I|x)$ and thus represent an alternative parametrization of the correlation space. Note that in the case of binary outcomes ($k=2$), $E(a_I|x_I)$ coincides with the usual definition of a correlation function. The definitions \eqref{eq:E} and \eqref{eq:E2} thus represent a possible generalisation of correlation function to more outcomes.

With the notation that we just introduced, a generic Bell inequality in the no-signalling space takes the form
\begin{equation}\label{incor}
\sum_{I,x_I,a_I} c(a_I,x_I) E(a_I|x_I) \leq c(0)
\end{equation}
where $c(a_I,x_I)$ are the coefficients of the inequality.
A property of the correlators $E(a_I|x_I)$ is that the white noise yields $E(a_I|x_I)=0$. The resistance to noise of a an inequality written in the form (\ref{incor}) is thus directly given by the ratio of the local bound to the violation.

A relabeling of parties, measurement settings, or measurement outcomes simply amounts to rearrange the order of the coefficients of the inequality. Note, however, that because of the normalization conditions (\ref{eq:normalisation}), the basis of the correlation space that we chose is overcomplete. The coefficients $c(a_I,x_I)$ are thus not uniquely defined: adding  $\sum_{a_i}E(a_I|x_I)=0$ for some $i\in I$ to the inequality (\ref{incor}) does not change the inequality itself, but does change its coefficients. To compare two inequalities, we must therefore ensure first that they are written in some standard way.

The freedom that we have in adding terms of the form $\lambda(i,a_{I\setminus i},x_I)\sum_{a_i}E(a_I|x_I)=0$ to (\ref{incor}) corresponds to define new coefficients for the inequality in the following manner
\be\label{eq:cprime}
c'(a_I,x_I)= c(a_I,x_I) +\sum_{i\in I} \lambda(i,a_{I\setminus i},x_I)\,.
\ee

We show now that requiring the inequality coefficients $c'(a_I,x_I)$ to satisfy the relation
\begin{equation}\label{eq:conditioncoeffs}
\sum_{a_i=0}^{k-1}c'(a_I,x_I)=0\quad \text{for all }i\in I\,.
\end{equation}
allows to define them uniquely.

\begin{proposition*}
There exist values of $\lambda(i,a_{I\setminus i},x_I)$ such that the newly defined coefficients $c'(a_I,x_I)$ satisfy relation \eqref{eq:conditioncoeffs}. Moreover for all such $\lambda$'s, the $c'(a_I,x_I)$ are the same, they are thus unique.
\end{proposition*}

\begin{proof}
Since equations \eqref{eq:cprime} and \eqref{eq:conditioncoeffs} apply independently on every subset $I$ of the parties, and on every inputs $x_I$, we omit these indices, writing for instance $c(a_I)$ instead of $c(a_I,x_I)$ to lighten the notation. Moreover, all sums on the outputs $a_i$ go from $0$ to $k-1$ and all sums on the parties $i$ run on $I$, so we also omit these bounds in the proof.

The existence of the $\lambda$'s can be shown by directly checking that the following formula is of the form \eqref{eq:cprime}, and satisfies \eqref{eq:conditioncoeffs}. Let
\begin{equation}\label{eq:cprimeformula}
c'(a_I)=\left[\left(\openone - \frac1k \sum_{a_1}\right)\circ\ldots\circ\left(\openone - \frac1k \sum_{a_n}\right)\right]c(a_I)
\end{equation}
where we used the notation $f + \sum_a f + \sum_b f = [\openone + \sum_a + \sum_b]f$ for any function $f$ and the $\circ$ composition satisfies $\openone \circ \openone = \openone$, $\openone \circ \sum_a = \sum_a \circ \openone = \sum_a$, $\sum_{a} \circ \sum_{b} = \sum_{a,b}$ and distributes over addition. To get this expression from \eqref{eq:cprime} one possible choice of lambdas is:
\begin{equation}\label{eq:lambdas}
\begin{split}
\lambda(1,a_{I\setminus 1}) &= -\frac1k \sum_{a_1} c(a_I)\\
\lambda(2,a_{I\setminus 2}) &= -\frac1k \sum_{a_2} c(a_I) + \left(\frac1k\right)^2 \sum_{a_1,a_2} c(a_I)\\
&\ldots
\end{split}
\end{equation}
and equation \eqref{eq:conditioncoeffs} is satisfied:
\begin{equation}
\begin{split}
\sum_{a_i}c'(a_I)&=\left[\sum_{a_i}\circ\left(\openone - \frac1k \sum_{a_1}\right)\circ\ldots\circ\left(\openone - \frac1k \sum_{a_n}\right)\right]c(a_I)\\
&=\left[\ldots\circ\left(\sum_{a_i} - \sum_{a_i}\right)\circ\ldots\right]c(a_I)=0.
\end{split}
\end{equation}

Now to show the unicity of the $c'$ coefficients, we notice that equation \eqref{eq:conditioncoeffs} is a non-homogeneous linear system of equations in the $\lambda$ variables, which we can write as:
\begin{equation}\label{eq:linsys}
\sum_{a_j}\sum_{i} \lambda(i,a_{I\setminus i}) = -\sum_{a_j} c(a_I)
\end{equation}
Every solution of this system can thus be written as $\lambda=\lambda_p+\lambda_v$ where $\lambda_p$ is a particular solution of the equation (as given by equation \eqref{eq:lambdas} for instance) and $\lambda_v$ is a solution of the homogeneous system, where the right-hand side of equation \eqref{eq:linsys} is replaced by zero. Thus every $c'(a_I)$ that satisfies equation \eqref{eq:conditioncoeffs} can be written as
\begin{equation}\label{eq:cpf}
c'(a_I) = c(a_I) + \sum_{i}\lambda_p(i,a_{I\setminus i}) + \sum_{i}\lambda_v(i,a_{I\setminus i}).
\end{equation}

Now we show that the last term of equation \eqref{eq:cpf} is zero for every solution $\lambda_v$ of the homogeneous counterpart of system \eqref{eq:linsys}, which implies that the coefficients $c'$ are uniquely defined. For this, consider the following expression:
\begin{equation}
Z=\left[\left(\openone-\frac1k\sum_{a_1}\right)\circ\ldots\circ\left(\openone-\frac1k\sum_{a_n}\right)\right]\lambda_v(i,a_{I\setminus i}).
\end{equation}
It is clearly zero, since it contains the term
\begin{equation}
\left[\left(\openone-\frac1k\sum_{a_i}\right)\right]\lambda_v(i,a_{I\setminus i})=\lambda_v(i,a_{I\setminus i})-\frac1k k \lambda_v(i,a_{I\setminus i})=0.
\end{equation}
On the other hand, we have that
\begin{align}
\sum_i Z &= \left[\left(\openone-\frac1k\sum_{a_1}\right)\circ\ldots\circ\left(\openone-\frac1k\sum_{a_n}\right)\right]\sum_i\lambda_v(i,a_{I\setminus i})\\
&=\left[\left(\openone-\frac1k\sum_{a_1}\right)\circ\ldots\circ\left(\openone-\frac1k\sum_{a_{n-1}}\right)\right]\sum_i\lambda_v(i,a_{I\setminus i})\nonumber\\
&\ \ \ -\left[\left(\openone-\frac1k\sum_{a_1}\right)\circ\ldots\circ\left(\openone-\frac1k\sum_{a_{n-1}}\right)\right]\frac1k\sum_{a_n}\sum_i\lambda_v(i,a_{I\setminus i})\label{eq:justbefore}\\
&=\left[\left(\openone-\frac1k\sum_{a_1}\right)\circ\ldots\circ\left(\openone-\frac1k\sum_{a_{n-1}}\right)\right]\sum_i\lambda_v(i,a_{I\setminus i})\\
\end{align}
where the second term in \eqref{eq:justbefore} vanishes by definition of $\lambda_v(i,a_{I\setminus i})$. Repeating iteratively the above step, we find eventually that
\begin{equation}
\sum_i Z = \sum_i\lambda_v(i,a_{I\setminus i}).
\end{equation}
This, combined with the fact that $Z=0$, implies the desired result.
\end{proof}

We showed that the coefficients of an inequality can be defined in a standard and unique way by requiring them to satisfy the constraints (\ref{eq:conditioncoeffs}). Now, since a relabeling of parties, measurement settings, or outcomes can only rearrange the coefficients $c'(a_I,x_I)$ without changing their value, the ordered lists of coefficients $c'(a_I,x_I)$ for $|I|=0,1,\ldots,n$ provide $n+1$ invariants for each equivalence class. For instance the local bound $c(0)$ is an (easily checkable) invariant.
In general requiring that two inequalities have their ordered list of coefficients identical does not guarantee that they are equivalent, but in the symmetric $(4,2,2)$ scenario for instance, all the $391$ classes of inequalities that we generated had different lists.

\section*{Appendix B: List of symmetric inequalities with two inputs and four outcomes.}
The following inequalities are given in the notation of \cite{Collins}.

\subsection*{Case with 3 outcomes for the first setting and 4 outcomes for the second one.}
\begin{narrow}{-2cm}{0cm}
\begin{equation}
\begin{split}
%
S_{(2,2,(3,4))}^{2}=\begin{array}{c||cc|ccc}
   & -1 & -1 &  0 &  0 &  0 \\
\hline \hline
-1 &  0 &  1 &  0 &  1 &  1 \\
-1 &  1 &  0 &  1 &  0 &  1 \\
\hline
 0 &  0 &  1 &  0 & -1 & -1 \\
 0 &  1 &  0 & -1 &  0 & -1 \\
 0 &  1 &  1 & -1 & -1 & -1 \\
\end{array} \leq 0\\
\end{split}
\end{equation}
\end{narrow}

\subsection*{Case with 4 outcomes for both settings.}
\begin{narrow}{ 0.6cm}{0cm}
\begin{equation}
\begin{aligned}
&S_{(2,2,4)}^{1}=\begin{array}{c||ccc|ccc}
&-1&-1&-1&0&0&0\\
\hline\hline
-1&0&0& 1& 1& 1& 1\\
-1&0& 1& 1& 1& 1&0\\
-1& 1& 1& 1& 1&0&0\\
\hline
0& 1& 1& 1&-1&-1&-1\\
0& 1& 1&0&-1&-1&0\\
0& 1&0&0&-1&0&0\\
\end{array}\leq0,
&
&S_{(2,2,4)}^{2}=\begin{array}{c||ccc|ccc}
&-1&-1&0&-1&0&0\\
\hline\hline
-1&0& 1&0& 1&0& 1\\
-1& 1&0&0& 1& 1&0\\
0&0&0&-1& 1&0&0\\
\hline
-1& 1& 1& 1& 1&0&0\\
0&0& 1&0&0&0&-1\\
0& 1&0&0&0&-1&0\\
\end{array}\leq0\\
&S_{(2,2,4)}^{3}=\begin{array}{c||ccc|ccc}
&-1&-1&0&-1&-1&0\\
\hline\hline
-1&0&0& 1& 1& 1&0\\
-1&0& 1&0&1&0&1\\
0& 1&0&0&-1&-1&0\\
\hline
-1& 1&1&-1&1&2&0\\
-1& 1&0&-1&2&2&0\\
0&0&1&0&0&0&-1\\
\end{array}\leq 0,
&
&S_{(2,2,4)}^{4}=\begin{array}{c||ccc|ccc}
&-1&-1&0&-1&-1&0\\
\hline\hline
-1&-1&1&-1&1&2&1\\
-1&1&1&0&1&0&1\\
0&-1&0&-1&1&1&0\\
\hline
-1&1&1&1& 1&0&0\\
-1&2&0&1&0& 1&-1\\
0&1&1&0&0&-1&-1\\
\end{array}\leq 0\\
&S_{(2,2,4)}^{5}=\begin{array}{c||ccc|ccc}
&-1&-1&-1&0&0&0\\
\hline\hline
-1&0& 1& 1&0&0& 1\\
-1& 1&0& 1&0& 1&0\\
-1& 1& 1&0& 1&0&0\\
\hline
0&0&0& 1&0&-1&-1\\
0&0& 1&0&-1&0&-1\\
0& 1&0&0&-1&-1&0\\
\end{array}\leq0,
&
&S_{(2,2,4)}^{6}=\begin{array}{c||ccc|ccc}
&-1&-1&-1&-1&0&0\\
\hline\hline
-1&0& 1& 1&0& 1&0\\
-1& 1& 1&0&1&0&1\\
-1& 1&0&-1&2&0&1\\
\hline
-1&0&1&2&1&-1&0\\
0&1&0&0&-1&0&-1\\
0&0&1&1&0&-1&-1\\
\end{array}\leq 0\\
&S_{(2,2,4)}^{7}=\begin{array}{c||ccc|ccc}
&-1&-1&-1&0&0&0\\
\hline\hline
-1&0& 1& 1&0&0& 1\\
-1& 1&0&0& 1& 1&0\\
-1& 1&0& 1& 1&0&0\\
\hline
0&0& 1& 1&-1&0&-1\\
0&0& 1&0&0&0&-1\\
0& 1&0&0&-1&-1&0\\
\end{array}\leq0,
&
&S_{(2,2,4)}^{8}=\begin{array}{c||ccc|ccc}
&-2&-1&-1&0&0&0\\
\hline\hline
-2& 2&0& 1&0& 1& 2\\
-1&0& 1& 1& 1&0&0\\
-1& 1& 1&0&0& 1&0\\
\hline
0&0& 1&0&-1&-1&0\\
0& 1&0& 1&-1&0&-1\\
0& 2&0&0&0&-1&-2\\
\end{array}\leq0
\end{aligned}
\end{equation}
\end{narrow}

\end{document}